\documentstyle[12pt]{article}
\date{}

\newtheorem{theorem}{Theorem}

\newcommand{\eth}{{\not\!\partial}}
\newcommand{\beth}{{\bar{\not\!\partial}}}

\newcommand{\lam}{{\lambda}}
\newcommand{\blam}{{\bar\lambda}}
\newcommand{\De}{{D'}}

\newcommand{\bA}{{\bar A}}
\newcommand{\al}{\alpha}
\newcommand{\bal}{{\bar\alpha}}
\newcommand{\be}{\beta}
\newcommand{\bbe}{{\bar\beta}}
\newcommand{\de}{\delta}
\newcommand{\bde}{{\bar\delta}}

\newcommand{\bm}{{\bar m}}

\newcommand{\sig}{{\sigma}}
\newcommand{\bga}{{\bar\gamma}}
\newcommand{\bsig}{{\bar\sigma}}
\newcommand{\btau}{{\bar\tau}}
\newcommand{\bmu}{{\bar\mu}}
\newcommand{\bnu}{{\bar\nu}}

\newcommand{\brho}{{\bar\rho}}
\newcommand{\bPsi}{{\bar\Psi}}
\newcommand{\tPsi}{{\tilde\Psi}}
\newcommand{\bze}{{\bar\zeta}}
\newcommand{\sY}{{{\ }_sY}}
\newcommand{\ppr}{{\frac{\partial}{\partial r}}}
\newcommand{\ppu}{{\frac{\partial}{\partial u}}}
\newcommand{\ppz}{{\frac{\partial}{\partial\zeta}}}
\newcommand{\ppbz}{{\frac{\partial}{\partial{\bar\zeta}}}}
\newcommand{\bee}{\begin{eqnarray}}
\newcommand{\ede}{\end{eqnarray}}
\begin{document}
\baselineskip=14 pt

\begin{center}
{\Large\bf On Newman-Penrose constants of stationary space-times }
\end{center}
\begin{center}
Xiaoning Wu${}^{a,}$\footnote{e-mail : wuxn@amss.ac.cn} and Yu
Shang${}^{a,b,}$\footnote{e-mail : amssshangyu@yahoo.com.cn }
\end{center}
\begin{center}
a. Institute of Applied Mathematics, \\
Academy of Mathematics and System Science,\\
Chinese Academy of Sciences,\\
 P.O.Box 2734, Beijing, China, 100080.\\
b. Graduate School of Chinese Academy of Science,\\
Beijing,  China, 100080
\end{center}
\begin{abstract}
We consider the general asymptotic expression of stationary
space-time. Using Killing equation, we reduce the dynamical freedom
of Einstein equation to the in-going gravitational wave $\Psi_0$.
The general form of this function can be got. With the help of
asymptotically algebraic special condition, we prove that all
Newman-Penrose constants vanish.
\end{abstract}



\section{Introduction}
In a classical work by Newman and Penrose \cite{NP68}, ten conserved
quantities are defined under some assumptions of asymptotic
flatness, that are called Newman-Penrose (N-P) constants, but the
physical meaning of these ten constants are still unclear. Much work
has been done. It is found that those set of constants play
important role in characterizing asymptotic flat space-time. On the
level of perturbative theory, Price \cite{Price} had found the N-P
constants of perturbative field strongly affected the long time
evolution of that filed. On the full non-linear theory, works by
Friedrich, K\'ann\'ar\cite{FK00} and Valiente Kroon\cite{Kr02} show
N-P constants contain the information that how the null infinity
${\cal I}^+$ touches spacial infinity $i^0$. In order to get more
understanding of the N-P constants, we need to consider some
examples for which we can really calculate the values of these
constants. People once guessed that algebraic special condition
might insure zero value of them, but Kinnersley and Walker\cite{Ki}
showed that this idea was incorrect. Some nice works have been done
to calculate the value of the N-P constants for some
examples\cite{LK00,DK02}. In this paper, we try to see what
restriction is brought by the stationary requirement. As we will
see, the stationary condition and an asymptotically algebraic
special condition will vanish all N-P constants.

This paper is organized as following : section II focuses on the
Taylor expansion of stationary space-time. With the help of Killing
equation, we reduce dynamical freedom of gravitational field into a
set of free-chosen constants. Detailed expression is given up to
order $o(r^{-6})$. Section III contains the main result of this
work. With algebraic special requirement, we prove all N-P constants
are zero. Finally in section IV some concluding remarks are made. In
Appendix, we give expression of some spin-weight harmonic functions
which is useful during the proof of the main theorem.

\section{The general asymptotic expansion of stationary space-time
in Bondi-Sachs' coordinates}

It is well-known that Bondi gauge choice\cite{PR86,UN62,St90} is a
powerful tool to study asymptotics of space-time. In many classical
works\cite{Bon,PR86,UN62}, the formal asymptotic expansion of
asymptotic flat space-time has been known in Bondi coordinates. With
this result, many important quantities, such as Bondi mass, Bondi
energy flux, can be expressed as integral on section of null
infinity. Let's focus on stationary space-times. Obviously, there is
no Bondi energy flux in such space-time, i.e. ${\dot\sig}^0=0$.
Further more, the super-translation freedom can help us to set
$\sigma^0$ to zero in stationary case \cite{NP68}. Submitting this
equation into the formal series, following the standard method in
\cite{PR86,UN62}, the Taylor expansions of stationary space-time
are:

1) Null tetrad
\begin{eqnarray}
l^a&=&\ppr\quad,\nonumber\\
n^a&=&\ppu+\left[-\frac{1}{2}-\frac{\Psi^0_2+\bPsi^0_2}{2r}+\frac{\beth\Psi^0_1+\eth\bPsi^0_1}{6r^2}\right.\nonumber\\
&&\quad\left.-\frac{\beth^2\Psi^0_0+\eth^2\bPsi^0_0}{24r^3}-\left(\frac{|\Psi^0_1|^2}{12}+\frac{\beth^2\Psi^1_0
+\eth^2\bPsi^1_0}{120}\right)r^{-4}+O(r^{-5})\right]\ppr\nonumber\\
&&+\left[\frac{1+\zeta\bze}{6\sqrt{2}r^3}\Psi^0_1
-\frac{1+\zeta\bze}{12\sqrt{2}r^4}\beth\Psi^0_0+O(r^{-5})\right]\ppz\nonumber\\
&&+\left[\frac{1+\zeta\bze}{6\sqrt{2}r^3}\bPsi^0_1
-\frac{1+\zeta\bze}{12\sqrt{2}r^4}\eth\bPsi^0_0+O(r^{-5})\right]\ppbz\quad,\nonumber\\
m^a&=&\left[-\frac{\Psi^0_1}{2r^2}+\frac{\beth\Psi^0_0}{6r^3}+\frac{\beth\Psi^1_0}{24r^4}+O(r^{-5})\right]\ppr\nonumber\\
&&+\left[\frac{1+\zeta\bze}{6\sqrt{2}r^4}\Psi^0_0+O(r^{-5})\right]\ppz
+\left[\frac{1+\zeta\bze}{\sqrt{2}r}+O(r^{-5})\right]\ppbz\quad.
\end{eqnarray}

2) N-P coefficients
\begin{eqnarray}
\rho&=&-\frac{1}{r}+O(r^{-6}),\nonumber\\
\sig&=&-\frac{\Psi^0_0}{2r^4}-\frac{\Psi^1_0}{3r^5}+O(r^{-6}),\nonumber\\
\al&=&\frac{\al^0}{r}-\frac{\bal^0\bPsi^0_0}{6r^4}+O(r^{-6}),\nonumber\\
\beta&=&-\frac{\bal^0}{r}-\frac{\Psi^0_1}{2r^3}+\frac{\al^0\Psi^0_0
+2\beth\Psi^0_0}{6r^4}+\frac{\beth\Psi^1_0}{8r^5}+O(r^{-6}),\nonumber\\
\tau&=&-\frac{\Psi^0_1}{2r^3}+\frac{\beth\Psi^0_0}{3r^4}+\frac{\beth\Psi^1_0}{8r^5}+O(r^{-6}),\nonumber\\
\lam&=&-\frac{\bPsi^0_0}{12r^4}-\frac{3\bPsi^0_0\Psi^0_2+\bPsi^1_0}{24r^5}+O(r^{-6}),\nonumber\\
\mu&=&-\frac{1}{2r}-\frac{\Psi^0_2}{r^2}+\frac{\beth\Psi^0_1}{2r^3}-\frac{\beth^2\Psi^0_0}{6r^4}
-\frac{-|\Psi^0_1|^2+\beth^2\Psi^1_0}{6r^5}+O(r^{-6}),\nonumber\\
\gamma&=&-\frac{\Psi^0_2}{2r^2}+\frac{2\beth\Psi^0_1+\al^0\Psi^0_1-\bal^0\bPsi^0_1}{6r^3}
-\left[\frac{1}{12}\left(\al^0\beth\Psi^0_0-\bal^0\eth\bPsi^0_0\right)+\frac{1}{8}\beth^2\Psi^0_0\right]r^{-4}\nonumber\\
&&+\left[-\frac{1}{40}\left(\al^0\beth\Psi^1_0-\bal^0\eth\bPsi^1_0\right)+\frac{1}{12}|\Psi^0_1|^2
-\frac{1}{30}\beth^2\Psi^1_0\right]r^{-5}+O(r^{-6}),\nonumber\\
\nu&=&-\left[\frac{1}{12}\bPsi^0_1+\frac{1}{6}\beth^2\Psi^0_1\right]r^{-3}
+\frac{1}{24}\left[\eth\bPsi^0_0+\beth^3\Psi^0_0\right]r^{-4}\nonumber\\
&&+\left[\frac{\bPsi^0_1\beth\Psi^0_1}{20}+\frac{\Psi^0_2\eth\bPsi^0_0}{15}+\frac{\eth\bPsi^1_0}{80}
-\frac{\bPsi^0_1\beth\Psi^0_1}{40}-\frac{\beth|\Psi^0_1|^2}{30}+\frac{\beth^3\Psi^1_0}{120}\right]r^{-5}+O(r^{-6}).
\end{eqnarray}

3) Weyl curvature
\begin{eqnarray}
&&\Psi_0=\qquad\ \ \qquad\ \ \frac{\Psi^0_0}{r^5}+\frac{\Psi^1_0}{r^6}+O(r^{-7}),\nonumber\\
&&\Psi_1=\qquad\ \ \frac{\Psi^0_1}{r^4}+\frac{\Psi^1_1}{r^5}+\frac{\Psi^2_1}{r^6}+O(r^{-7}),\nonumber\\
&&\Psi_2=\frac{\Psi^0_2}{r^3}+\frac{\Psi^1_2}{r^4}+\frac{\Psi^2_2}{r^5}+\frac{\Psi^3_2}{r^6}+O(r^{-7}),\\
&&\Psi_3=\qquad\ \ \frac{\Psi^2_3}{r^4}+\frac{\Psi^3_3}{r^5}+\frac{\Psi^4_3}{r^6}+O(r^{-7}),\nonumber\\
&&\Psi_4=\qquad\ \ \qquad\ \
\frac{\Psi^4_4}{r^5}+\frac{\Psi^5_4}{r^6}+O(r^{-7}).\nonumber
\end{eqnarray}
where
\begin{eqnarray}
&&\Psi^1_1=-\beth\Psi^0_0,\
\Psi^2_1=-\frac{1}{2}\beth\Psi^1_0,\nonumber\\
&&\Psi^0_2=-M,\ \Psi^1_2=-\frac{1}{4}\beth\Psi^0_1,\
\Psi^2_2=\frac{1}{2}\beth^2\Psi^0_0,\
\Psi^3_2=[-\frac{2}{3}|\Psi^0_1|^2+\frac{1}{6}\beth^2\Psi^1_0],\nonumber\\
&&\Psi^2_3=\frac{1}{2}\beth^2\Psi^0_1, \
\Psi^3_3=[-\frac{1}{2}\bPsi^0_1\Psi^0_2-\frac{1}{6}\beth^3\Psi^0_0],
\
\Psi^4_3=\frac{1}{8}\bPsi^0_1\beth\Psi^0_1+\frac{1}{6}\beth|\Psi^0_1|^2-\frac{1}{24}\beth^3\Psi^1_0,\nonumber\\
&&\Psi^4_4=\frac{1}{24}\beth^4\Psi^0_0,\
\Psi^5_4=\frac{9}{40}\bPsi^0_1\beth^2\Psi^0_1-\frac{1}{30}\beth^3|\Psi^0_1|^2+\frac{1}{120}\beth^4\Psi^1_0
-\frac{1}{20}\Psi^0_2\bPsi^0_0.\nonumber
\end{eqnarray}
Based on the results of characteristic initial value
problem\cite{Fr81,Fr82,Ka96,Ni06}, we know the dynamical freedom of
asymptotic space-time are characterized by $\{\sig^0, \Psi_1,
Re(\Psi_2)\}$ on ${\cal I}^+$ and $\Psi_4$ on $N$, where $N$ is an
out-going light cone of the space-time. Obviously, the stationary
condition has eliminated the freedom of news function. In this
section, we will use Killing equation to reduce other dynamical
freedom and get a general asymptotic extension of stationary
space-time. As we have done in \cite{Wu06}, we choose the standard
Bondi-Sachs' coordinates, use the standard Bondi null tetrad and
gauge choice\cite{PR86,UN62}, express the time-like Killing vector
as
\begin{eqnarray}
t^a=Tl^a+Rn^a+\bA m^a+A\bm^a.\nonumber
\end{eqnarray}
The Killing equations are
\begin{eqnarray}
&&-DR=0,\label{a1}\\
&&-DT-\De R+(\gamma+\bga)R+\btau
A+\tau\bA=0,\label{a2}\\
&&DA-\de R+\tau R+\brho
A+\sig\bA=0,\label{a3}\\
&&-\De T-(\gamma+\bga)T-\nu A-\bnu\bA=0,\label{a4}\\
&&-\tau T+\bnu R+\De A+(\bga-\gamma)A-\de T-\tau T-\mu
A-\blam\bA=0,\label{a5}\\
&&-\sig T+\blam R+\de A+(\bal-\be)A=0,\label{a6}\\
&&-\rho T+\mu R+\de\bA-(\bal-\be)\bA-\brho T+\bmu R+\bde
A-(\al-\bbe)A=0.\label{a7}
\end{eqnarray}
Similar as the analysis in \cite{Wu06}, we can normalize function
$R$ to 1. Assuming the asymptotic behavior of $A$ and $T$ as
\begin{eqnarray}
T&=&T^0+\frac{T^1}{r}+\cdots,\nonumber\\
A&=&A^0+\frac{A^1}{r}+\cdots,\label{TA}
\end{eqnarray}
we can solve Killing equations order by order. The stationary
condition implies ${\dot\sig}^0=0$. Consequently, we have
$\Psi^0_3=0$ and $\Psi^0_4=0$ because of the Bianchi identity. The
first order Killing equations are
\begin{eqnarray}
&&-A^0=0,\\
&&-{\dot T}^1+\Psi^0_3A^0+\bPsi^0_3\bA^0=0,\\
&&-\bPsi^0_3+{\dot A}^1+\frac{1}{2}A^0-{\dot\bsig}^0A^0=0,\\
&&{\dot\bsig}^0+\de_0A^0+2\bal^0A^0=0,\\
&&2T^0-1=0,
\end{eqnarray}
which implies ${\dot T}^1=0$, ${\dot A}^1=0$.

The second order Killing equations are
\begin{eqnarray}
&&T^1-\frac{1}{2}(\Psi^0_2+\bPsi^0_2)=0,\\
&&-2A^1=0,\\
&&{\dot T}^2=0,\\
&&\frac{1}{2}\eth\Psi^0_2+{\dot A}^2-\de_0T^1=0,\\
&&\frac{1}{2}\sig^0=0,\\
&&2T^1-\sig^0{\dot\bsig}^0-\bsig^0{\dot\sig}^0-\Psi_2^0-\bPsi^0_2=0,
\end{eqnarray}
where $\eth f=(\de_0+2s\bal^0)f$. From these equations, we know
$\sig^0=0$, $A^1=0$, ${\dot T}^2=0$,
$T^1=\frac{1}{2}(\Psi^0_2+\bPsi^0_2)$, $\Psi^0_2=\bPsi^0_2$,
${\dot\Psi}^0_2=0$, ${\dot
A}^2=-\frac{1}{2}\eth\Psi^0_2+\frac{1}{2}\de_0(\Psi^0_2+\bPsi^0_2)$.

The third order equations are
\begin{eqnarray}
&&2T^2+\frac{1}{3}(\beth\Psi^0_1+\eth\bPsi^0_1)=0,\label{23}\\
&&-3A^2-\frac{1}{2}\Psi^0_1=0,\label{33}\\
&&{\dot T}^3=0,\label{43}\\
&&\frac{1}{2}\Psi^0_1+\bnu^3+{\dot
A}^3+\frac{3}{2}A^2-\de_0T^2=0,\label{53}\\
&&\de_0A^2+2\bal^0A^2=0,\label{63}\\
&&2T^2+\frac{1}{2}\beth\Psi^0_1+\frac{1}{2}\eth\bPsi^0_1+\de_0\bA^2-2\bal^0\bA^2+\bde_0A^2-2\al^0A^2=0.
\end{eqnarray}
Eq.(\ref{33}),(\ref{63}) imply
\begin{eqnarray}
\eth\Psi^0_1=0.
\end{eqnarray}
Because the spin-weight of $\Psi^0_1$ is 1, we can express it as a
linear combination of spin-weight harmonics, i.e.
\begin{eqnarray}
\Psi^0_1=\sum^1_{m=-1}c_m(u){\ }_1Y_{1,m}.
\end{eqnarray}
The action of $\eth$ and $\beth$ on spin-weight harmonics are
\cite{PR86}
\begin{eqnarray}
\eth\sY_{lm}&=&-\sqrt{\frac{(l+s+1)(l-s)}{2}}{\ }_{s+1}Y_{lm},\nonumber\\
\beth\sY_{lm}&=&\ \ \sqrt{\frac{(l-s+1)(l+s)}{2}}{\
}_{s-1}Y_{lm}.\label{swh}
\end{eqnarray}
Combining Eq.(\ref{23}) with ${\dot T}^2=0$, we get
\begin{eqnarray}
0&=&\sum^1_{m=-1}\left({\dot c}_m\ Y_{1,m}+{\dot{\bar c}}_m\ {\bar
Y}_{1,m}\right)\nonumber\\
&=&({\dot c}_{-1}-{\dot{\bar c}}_1)Y_{1,-1}+({\dot c}_{0}+{\dot{\bar
c}}_0)Y_{1,0}+({\dot c}_{1}-{\dot{\bar c}}_{-1})Y_{1,1}.\nonumber
\end{eqnarray}
This means ${\dot c}_{-1}-{\dot{\bar c}}_1=0$ and ${\dot
c}_{0}+{\dot{\bar c}}_0=0$. Bianchi identity
${\dot\Psi}^0_1-\eth\Psi^0_2=0$ gives
\begin{eqnarray}
\Psi^0_2=-{\dot c}_{-1}\ Y_{1,-1}-{\dot c}_0\ Y_{1,0}-{\dot c}_1\
Y_{1,1}+C.\nonumber
\end{eqnarray}
The stationary condition insures ${\dot\sig}^0=0$, which implies
$\Psi^0_2=\bPsi^0_2$. This leads to
\begin{eqnarray}
{\dot c}_{-1}+{\dot{\bar c}}_1=0,\quad{\dot c}_0-{\dot{\bar
c}}_0=0\quad and\quad C={\bar C}.\nonumber
\end{eqnarray}
It is easy to see that ${\dot c}_m=0$ and $\Psi^0_2=C$, so we have
\begin{eqnarray}
\Psi^0_1&=&\sum^1_{m=-1}B_m{\ }_1Y_{1,m},\nonumber\\
\Psi^0_2&=&C.\label{Psi1}
\end{eqnarray}
The Komar integral shows that $-C$ is just the Bondi mass of this
space-time and $\{B_m\}$ can be understood as the components of
angular momentum. Here we notice that stationary condition helps us
to restrict the form of initial data $\Psi^0_1$ and $Re(\Psi^0_2)$.
The only freedom left is just $\Psi_0$ which carries information of
in-going gravitational wave.

In order to get information of $\Psi_0$, we need to consider higher
order behavior of Killing equations. The forth order Killing
equations are
\begin{eqnarray}
&&3T^3+(\gamma^4+\bga^4)=0,\label{24}\\
&&4A^3=\frac{1}{3}\beth\Psi^0_0,\label{34}\\
&&{\dot T}^4+\frac{1}{3}(\beth\Psi^0_1+\eth\bPsi^0_1)(\Psi^0_2+\bPsi^0_2)=0,\label{44}\\
&&\frac{1}{2}\Psi^0_1T^1-\frac{1}{3}\beth\Psi^0_0+\bnu^4+{\dot
A}^4+(\Psi^0_2+\bPsi^0_2)A^2+\frac{3}{2}A^3-\de_0 T^3+\Psi^0_2A^2+\frac{1}{2}A^3=0,\label{54}\\
&&\frac{1}{4}\Psi^0_0+\blam^4+\eth A^3=0,\label{64}\\
&&2T^3+\mu^4+\bmu^4+\eth\bA^3+\beth A^3=0.\label{74}
\end{eqnarray}
Eq.(\ref{34}),(\ref{64}) imply \bee \Psi^0_0=\sum^2_{m=-2}A_m(u){\
}_2Y_{2,m},\label{Psi00} \ede Eq.(\ref{53}),(\ref{34}) insure that
$\Psi^0_0$ is independent of $u$.

Our aim is to consider the Newman-Penrose constants, which are
contained in coefficient $\Psi^1_0$. So the fifth order of Killing
equations is needed.
\begin{eqnarray}
&&4T^4+(\gamma^5+\bga^5)-\frac{1}{2}\bPsi^0_1A^2-\frac{1}{2}\Psi^0_1\bA^2=0,\label{25}\\
&&A^4=\frac{1}{5}\tau^5,\label{35}\\
&&-\frac{1}{2}\sig^5+\blam^5+\frac{1}{2}\Psi^0_0T^1+\frac{3}{2}\Psi^0_1A^2+\eth A^4=0,\label{65}\\
&&-2\rho^5+2T^4+(\mu^5+\bmu^5)+\frac{3}{2}\Psi^0_1\bA^2+\frac{3}{2}\bPsi^0_1A^2
+\eth\bA^4+\beth A^4=0,\label{75}
\end{eqnarray}
Eq.(\ref{35}),(\ref{65}) give
\begin{eqnarray}
\eth\beth\Psi^1_0+5\Psi^1_0=10(\Psi^0_1)^2-15\Psi^0_0\Psi^0_2.\label{Psi01}
\end{eqnarray}
Equation above is the key one which helps us to control the N-P
constants of stationary space-time in next section. It has been
noticed that each order of Killing equation helps us to control the
Taylor coefficients of the same order, just like
Eq.(\ref{Psi1}),(\ref{Psi00}),(\ref{Psi01}). Similarly, higher order
Killing equations are also with the same property. The general form
should be
\begin{eqnarray}
\eth\beth{\Psi}^{k}_0+\frac{(k+4)(k+1)}{2}{\Psi}^{k}_0=f.
\end{eqnarray}
where $f$ is known function\cite{Wu06}. The uncertainty comes from
the homogeneous part of above equations. There are some free
constants in each order. These arbitrary constants should be closely
related to the famous Geroch-Hansen multi-pole moments
\cite{Ger70,Han74,Fr06}.

\section{N-P constants of stationary, algebraic special space-times}
In last section, we get the Taylor expansions of general asymptotic
flat, stationary space-times up to order $o(r^{-6})$ which is enough
for us to calculate the N-P constants. The Killing equation also
helps us to restrict the dynamical freedom. We have found the
information of Bondi mass is contained in $\Psi^0_2$, angular
momentum is in $\Psi^0_1$. By definition\cite{PR86}, N-P constants
are
\begin{eqnarray}
G_m=\int_{S_{\infty}}{}_2Y_{2,m}\Psi^1_0 dS.
\end{eqnarray}
Back to eq.(\ref{Psi01}), we know
\begin{eqnarray}
\eth\beth\Psi^1_0+5\Psi^1_0=10(\Psi^0_1)^2-15\Psi^0_0\Psi^0_2.\label{Psi011}
\end{eqnarray}
This is an inhomogeneous linear equation. The associated homogeneous
equation is
\begin{eqnarray}
\eth\beth\Psi+5\Psi=0.
\end{eqnarray}
Keeping Eq.(\ref{swh}) in mind, we can get the general solution of
above equation:
$$\Psi=\sum^3_{m=-3}D_m{\ }_2Y_{3,m}\qquad.$$
Let $3\tPsi^1_0=10(\Psi^0_1)^2-15\Psi^0_2\Psi^0_0$, with the help of
Eq.(\ref{Psi1}) and (\ref{Psi00}), we know
\begin{eqnarray}
\eth\beth\tPsi^1_0+5\tPsi^1_0=10(\Psi^0_1)^2-15\Psi^0_0\Psi^0_2,
\end{eqnarray}
then the general solution of Eq.(\ref{Psi011}) is
\begin{eqnarray}
\Psi^1_0=\tPsi^1_0+\sum^3_{m=-3}D_m{\ }_2Y_{3,m}\qquad.
\end{eqnarray}
Because of the definition, it is clear that only $\tPsi^1_0$
contributes to the N-P constants. Combining above equation with
Eq.(\ref{Psi1}),(\ref{Psi00}), we have
\begin{eqnarray}
\tPsi^1_0&=&\frac{10}{3}(\Psi^0_1)^2-5\Psi^0_2\Psi^0_0\nonumber\\
&=&\frac{5}{3}\times[2(\sum^1_{m=-1}B_m{\ }_1Y_{1,m})^2+3M(\sum^2_{m=-2}A_m{\ }_2Y_{2,m})]\nonumber\\
&=&\frac{10}{3}[B_{-1}{\ }_{1}Y_{1,-1}+B_0{\
}_{1}Y_{1,0}+B_1{\
}_{-1}Y_{1,1}]^2\nonumber\\
&&+5M[A_{-2}{\ }_{2}Y_{2,-2}+A_{-1}{\ }_{2}Y_{2,-1}+A_0{\
}_{2}Y_{2,0}+A_1{\ }_{2}Y_{2,-1}+A_2{\
}_{2}Y_{2,2}]\nonumber\\
&=&\frac{10}{3}[(B_{-1})^2({\ }_{1}Y_{1,-1})^2+(B_0)^2({\
}_{1}Y_{1,0})^2+(B_1)^2({\ }_{1}Y_{1,1})^2\nonumber\\
&&\qquad +2B_{-1}B_0{\ }_{1}Y_{1,-1}{\ }_{1}Y_{1,0}+2B_0B_1{\
}_{1}Y_{1,0}{\ }_{1}Y_{1,1}+B_{-1}B_1{\ }_{1}Y_{1,-1}{\
}_{1}Y_{1,1}]\nonumber\\
&&+5M[A_{-2}{\ }_{2}Y_{2,-2}+A_{-1}{\ }_{2}Y_{2,-1}+A_0{\
}_{2}Y_{2,0}+A_1{\ }_{2}Y_{2,-1}+A_2{\ }_{2}Y_{2,2}]\nonumber\\
&=&\frac{5}{3}\left\{[2(B_{-1})^2\sqrt{\frac{3}{40\pi}}+3MA_{-2}]{\
}_{2}Y_{2,-2}+[4B_{-1}B_0\sqrt{\frac{3}{80\pi}}+3MA_{-1}]{\
}_{2}Y_{2,-1}\right.\nonumber\\
&&\qquad+[4(B_0)^2\sqrt{\frac{1}{80\pi}}+4B_{-1}B_1\sqrt{\frac{1}{80\pi}}+3MA_0]{\
}_{2}Y_{2,0}\nonumber\\
&&\qquad\left.+[4B_0B_1\sqrt{\frac{3}{80\pi}}+3MA_1]{\
}_{2}Y_{2,1}+[2(B_1)^2\sqrt{\frac{3}{40\pi}}+3MA_2]{\
}_{2}Y_{2,2}\right\}\nonumber\\
&=&\sum^2_{m=-2}G_m{\ }_{2}Y_{2,m}.\label{np0}
\end{eqnarray}
By definition, $\{G_m\}$ are just N-P constants. In general, it's
easy to understand that N-P constants may not be zero because there
are free constants $\{B_m\}$ and $\{A_m\}$ which are out of our
control.

Now, we consider the algebraic special case. The algebraic special
condition is \cite{Kr80}
\begin{eqnarray}
 &&I^3=27J^2,\label{as}\\
&&I=\Psi_0\Psi_4-4\Psi_1\Psi_3+3(\Psi_2)^2,\nonumber\\
&&J=\Psi_4\Psi_2\Psi_0+2\Psi_3\Psi_2\Psi_1-(\Psi_2)^3-(\Psi_3)^2\Psi_0-(\Psi_1)^2\Psi_4.\nonumber
\end{eqnarray}
Submitting the asymptotic extension of Weyl curvature into above
equations, we get the first non-zero equation as
\begin{eqnarray}
0&=&36(\Psi^0_1)^2(\Psi^0_2)^2(\Psi^2_3)^2-54(\Psi^0_2)^3(\Psi^2_3)^2\Psi^0_0
-54(\Psi^0_1)^2(\Psi^0_2)^3\Psi^4_4+81(\Psi^0_2)^4\Psi^0_0\Psi^4_4\nonumber\\
&=&9[(\Psi^0_1)^2(\beth^2\Psi^0_1)^2+\frac{3M}{2}(\beth^2\Psi^0_1)^2\Psi^0_0
+6M(\Psi^0_1)^2\frac{1}{24}\beth^4\Psi^0_0+\frac{9}{24}M^2\Psi^0_0\beth^4\Psi^0_0]\nonumber\\
&=&9[(\sum^1_{m=-1}B_m{\ }_1Y_{1,m})^2(\sum^1_{m=-1}B_m{\
}_{-1}Y_{1,m})^2+\frac{3M}{2}(\sum^1_{m=-1}B_m{\
}_{-1}Y_{1,m})^2(\sum^2_{m=-2}A_m{\ }_2Y_{2,m})\nonumber\\
&&+\frac{3M}{2}(\sum^1_{m=-1}B_m{\ }_1Y_{1,m})^2(\sum^2_{m=-2}A_m{\
}_{-2}Y_{2,m})+\frac{9M^2}{4}(\sum^2_{m=-2}A_m{\
}_2Y_{2,m})(\sum^2_{m=-2}A_m{\ }_{-2}Y_{2,m})]\nonumber\\
&=&\frac{9}{4}\times[2(\sum^1_{m=-1}B_m{\
}_1Y_{1,m})^2+3M(\sum^2_{m=-2}A_m{\
}_2Y_{2,m})]\nonumber\\
&&\times[2(\sum^1_{m=-1}B_m{\ }_{-1}Y_{1,m})^2+3M(\sum^2_{m=-2}A_m{\
}_{-2}Y_{2,m})],
\end{eqnarray}
which means
\begin{eqnarray}
&&2(\sum^1_{m=-1}B_m{\ }_1Y_{1,m})^2+3M(\sum^2_{m=-2}A_m{\
}_2Y_{2,m})=0\label{c1}\\
&&or\nonumber\\
&&2(\sum^1_{m=-1}B_m{\ }_{-1}Y_{1,m})^2+3M(\sum^2_{m=-2}A_m{\
}_{-2}Y_{2,m})=0.\label{c2}
\end{eqnarray}
Comparing Eq.(\ref{c1}) with the second line of eq.(\ref{np0}), it
is easy to see that N-P constants vanish if Eq.(\ref{c1}) holds. If
Eq.(\ref{c2}) holds, we have
\begin{eqnarray}
0&=&2(\sum^1_{m=-1}B_m{\ }_{-1}Y_{1,m})^2+3M(\sum^2_{m=-2}A_m{\
}_{-2}Y_{2,m})\nonumber\\
&=&2[B^2_1({\ }_{-1}Y_{1,1})^2+B^2_0({\ }_{-1}Y_{1,0})^2+B^2_{-1}({\
}_{-1}Y_{1,-1})^2\nonumber\\
&&\quad+2B_1B_0{\ }_{-1}Y_{1,1}{\ }_{-1}Y_{1,0}+2B_1B_{-1}{\
}_{-1}Y_{1,1}{\ }_{-1}Y_{1,-1}+2B_0B_{-1}{\ }_{-1}Y_{1,0}{\
}_{-1}Y_{1,-1}]\nonumber\\
&&+3MA_2{\ }_{-2}Y_{2,2}+3MA_1{\ }_{-2}Y_{2,1}+3MA_0{\
}_{-2}Y_{2,0}+3MA_{-1}{\ }_{-2}Y_{2,-1}+3MA_{-2}{\
}_{-2}Y_{2,-2}\nonumber\\
&=&\left[2\sqrt{\frac{3}{40\pi}}B^2_1+3MA_2\right]{}_{-2}Y_{2,2}
+\left[4\sqrt{\frac{3}{80\pi}}B_1B_0+3MA_1\right]{}_{-2}Y_{2,1}\nonumber\\
&&+\left[\frac{4}{\sqrt{80\pi}}(B^2_0+B_1B_{-1})+3MA_0\right]{}_{-2}Y_{2,0}\nonumber\\
&&+\left[4B_{-1}B_0\sqrt{\frac{3}{80\pi}}+3MA_{-1}\right]{
}_{-2}Y_{2,-1}+\left[2(B_{-1})^2\sqrt{\frac{3}{40\pi}}+3MA_{-2}\right]{
}_{-2}Y_{2,-2}.
\end{eqnarray}
Comparing above equation with the fifth line of Eq.(\ref{np0}),
Eq.(\ref{c2}) also implies zero value of the N-P constants, so we
have proved following result
\begin{theorem}
All N-P constants of asymptotic flat, stationary, algebraic special
space-time vanish.
\end{theorem}
Remark : What needs to emphasis is that Kerr space-time is an
important special case of our theorem, i.e. all N-P constants of
Kerr space-time are zero. Based on our knowledge, the value of the
N-P constants of Kerr space-time is unknown. The value of N-P
constants of Kerr space-time are also got by us with other
methods\cite{Wu06,Bai06}.

\section{Discussion}
In last section, we prove a theorem that asymptotic flat,
stationary, algebraic special condition vanishes all N-P constants.
In fact, our result is a bit stronger than above theorem because
only the leading order of the algebraic special condition is needed.
Maybe we can call such kind of space-time as asymptotically
algebraic special space-time. This tells us that algebraic special
condition may not be a very suitable choice to control the N-P
constants. A natural question is whether the stationary condition is
enough to vanish all N-P constants in vacuum case. As we have seen,
the stationary condition has strongly restricted the freedom of
gravitational field. If this idea is right, it would give a physical
understanding for N-P constants. Although we still haven't very
strong evidence, some primary calculation implies this idea maybe
right. This will be discussed in future works.

\section*{Acknowledgement}
This work is supported by National Science Foundation of China
under Grant No. 10375087 and K.C.Wong Education Foundation, Hong
Kong. Authors would like to thank Prof.Y.K.Lau, Dr.Z.J.Cao, S.Bai
and X.F.Gong for their helpful discussions. X.Wu also wants to
thank Prof.H.Friedrich and Dr.J.A.Valiente Kroon for their helpful
discussion.

\section*{Appendix}
Some spin-weight harmonics
\begin{eqnarray}
Y_{0,0}&=&\frac{1}{\sqrt{4\pi}};\nonumber\\
{}_1Y_{1,1}&=&\sqrt{\frac{3}{16\pi}}(\cos\theta+1)e^{i\phi},\nonumber\\
{}_1Y_{1,0}&=&\sqrt{\frac{3}{8\pi}}\sin\theta,\nonumber\\
{}_1Y_{1,-1}&=&\sqrt{\frac{3}{16\pi}}(1-\cos\theta)e^{-i\phi};\nonumber\\
Y_{1,1}&=&-\sqrt{\frac{3}{8\pi}}\sin\theta e^{i\phi},\nonumber\\
Y_{1,0}&=&\sqrt{\frac{3}{4\pi}}\cos\theta,\nonumber\\
Y_{1,-1}&=&\sqrt{\frac{3}{8\pi}}\sin\theta e^{-i\phi};\nonumber\\
{}_{-1}Y_{1,1}&=&\sqrt{\frac{3}{16\pi}}(1-\cos\theta)e^{i\phi},\nonumber\\
{}_{-1}Y_{1,0}&=&-\sqrt{\frac{3}{8\pi}}\sin\theta,\nonumber\\
{}_{-1}Y_{1,-1}&=&\sqrt{\frac{3}{16\pi}}(1+\cos\theta)e^{-i\phi};\nonumber\\
{}_{2}Y_{2,2}&=&3\sqrt{\frac{5}{96\pi}}(1+\cos\theta)^2e^{2i\phi},\nonumber\\
{}_{2}Y_{2,1}&=&3\sqrt{\frac{5}{24\pi}}\sin\theta(1+\cos\theta)e^{i\phi},\nonumber\\
{}_{2}Y_{2,0}&=&3\sqrt{\frac{5}{16\pi}}\sin^2\theta,\nonumber\\
{}_{2}Y_{2,-1}&=&3\sqrt{\frac{5}{24\pi}}\sin\theta(1-\cos\theta)e^{-i\phi},\nonumber\\
{}_{2}Y_{2,-2}&=&3\sqrt{\frac{5}{96\pi}}(1-\cos\theta)^2e^{-2i\phi};\nonumber\\
{}_{1}Y_{2,2}&=&-3\sqrt{\frac{5}{24\pi}}\sin\theta(1+\cos\theta)e^{2i\phi},\nonumber\\
{}_{1}Y_{2,1}&=&3\sqrt{\frac{5}{24\pi}}(2\cos\theta-1)(1+\cos\theta)e^{i\phi},\nonumber\\
{}_{1}Y_{2,0}&=&3\sqrt{\frac{5}{4\pi}}\sin\theta\cos\theta,\nonumber\\
{}_{1}Y_{2,-1}&=&3\sqrt{\frac{5}{24\pi}}(2\cos\theta+1)(1-\cos\theta)e^{-i\phi},\nonumber\\
{}_{1}Y_{2,-2}&=&3\sqrt{\frac{5}{24\pi}}\sin\theta(1-\cos\theta)e^{-2i\phi};\nonumber\\
{}_{}Y_{2,2}&=&3\sqrt{\frac{5}{16\pi}}\sin^2\theta e^{2i\phi},\nonumber\\
{}_{}Y_{2,1}&=&-6\sqrt{\frac{5}{16\pi}}\sin\theta\cos\theta
e^{i\phi},\nonumber\\
{}_{}Y_{2,0}&=&\sqrt{\frac{5}{24\pi}}(3\cos^2\theta-1),\nonumber\\
{}_{}Y_{2,-1}&=&6\sqrt{\frac{5}{16\pi}}\sin\theta\cos\theta
e^{-i\phi},\nonumber\\
{}_{}Y_{2,-2}&=&3\sqrt{\frac{5}{16\pi}}\sin^2\theta e^{-2i\phi};\nonumber\\
{}_{-1}Y_{2,2}&=&-3\sqrt{\frac{5}{24\pi}}\sin\theta(1-\cos\theta)e^{2i\phi},\nonumber\\
{}_{-1}Y_{2,1}&=&3\sqrt{\frac{5}{24\pi}}(2\cos\theta+1)(1-\cos\theta)e^{i\phi},\nonumber\\
{}_{-1}Y_{2,0}&=&-\sqrt{\frac{5}{4\pi}}\sin\theta\cos\theta,\nonumber\\
{}_{-1}Y_{2,-1}&=&3\sqrt{\frac{5}{24\pi}}(2\cos\theta-1)(1+\cos\theta)e^{-i\phi},\nonumber\\
{}_{-1}Y_{2,-2}&=&3\sqrt{\frac{5}{24\pi}}\sin\theta(1+\cos\theta)e^{-2i\phi};\nonumber\\
{}_{-2}Y_{2,2}&=&3\sqrt{\frac{5}{96\pi}}(1-\cos\theta)^2e^{2i\phi},\nonumber\\
{}_{-2}Y_{2,1}&=&-3\sqrt{\frac{5}{24\pi}}\sin\theta(1-\cos\theta)e^{i\phi},\nonumber\\
{}_{-2}Y_{2,0}&=&3\sqrt{\frac{5}{16\pi}}\sin^2\theta,\nonumber\\
{}_{-2}Y_{2,-1}&=&-3\sqrt{\frac{5}{24\pi}}\sin\theta(1+\cos\theta)e^{-i\phi},\nonumber\\
{}_{-2}Y_{2,-2}&=&3\sqrt{\frac{5}{96\pi}}(1+\cos\theta)^2e^{-2i\phi}.\nonumber
\end{eqnarray}


\begin{thebibliography}{99}
\bibitem{NP68} E.~T.~Newman and R.~Penrose, Proc. Roy. Soc. Lond.
{\bf A} {\bf 305} (1968) 175.

\bibitem{Price}R.~H.~Price, Phys. Rev. {\bf D} {\bf 5} (1972)
2419.

\bibitem{FK00} H. Friedrich and J. K\'ann\'ar, J. Math. Phys. {\bf 41} (2000)
2195.

\bibitem{Kr02}J.A. Valiente Kroon, Class. Quant. Grav. {\bf 20} (2003)
L53.

\bibitem{Ki}W.~Kinnersely and M.~Walker, Phys. Rev. {\bf D} {\bf
2} (1970) 1359.

\bibitem{LK00}R.~Lazkoz and J.~A.~Valiente Kroon, Phys. Rev. {\bf
D}{\bf 62} (2000) 084033.

\bibitem{DK02}S.~Dain and J.~A.~Valiente Kroon, Class. Quant. Grav., {\bf 19}(2002)811.

\bibitem{Bon}H.~Bondi, M.~G.~J.~van der Burg and A.W.K.Metzner,
Proc. Roy. Soc. Lond. {\bf A} {\bf 269} (1962) 21.

\bibitem{PR86}R.~Penrose and R.~Rindler, {\it Spinors and
Space-Time} Vol.I and II, Cambridge University Press, 1986.

\bibitem{UN62}E.~T.~Newman and T.~W.~J.~Unti, J. Math. Phys. {\bf 3}
(1962 ) 891.

\bibitem{St90}J.~Stewart, {\it Advanced General Relativity},
Cambridge University Press, 1990.

\bibitem{Fr81}H.~Friedrich, Proc. Roy. Soc. Lond. {\bf A} {\bf 378}
(1981) 169-184, 401-421

\bibitem{Fr82}H.~Friedrich, Proc. Roy. Soc. Lond. {\bf A} {\bf
381} 361-371.

\bibitem{Ka96} J.~K\'ann\'ar, Proc. Roy. Soc. Lond. {\bf A} {\bf 452}
(1996) 945.

\bibitem{Ni06}F.~Nicol\'o, ``A local characteristic problem for the Einstein vacuum equations", gr-qc/0603118.

\bibitem{Ger70}R.~Geroch, J. Math. Phys. {\bf 11} (1970) 1955,
2580.

\bibitem{Han74}R.~Hansen, J. Math. Phys. {\bf 15} (1974) 46.

\bibitem{Fr06}H.~Friedrich, ``Static vacuum solutions from
convergent null dadta expensions at space-like infinity",
gr-qc/0606133.

\bibitem{Kr80}D.~Kramer, H.~Stephani, E.~Herlt and M.~MacCallum,
{\it Exact Solutions of Einstein's Field Equations}, Cambridge
University Press, 1980.

\bibitem{Wu06}X.~Wu and S.~Bai, ``On local uniqueness of Kerr space-time", in preparing.

\bibitem{Bai06}S.~Bai et.al., ``Bondi-Sachs formulism of Kerr metric",
submit to Phys. Rev. {\bf D}.

\end{thebibliography}
\end{document}